\documentclass[reprint,
superscriptaddress,
 amsmath,amssymb,
 aps,
pra,
]{revtex4-2}

\usepackage{caption}
\usepackage{subcaption}
\usepackage{graphicx}
\usepackage{dcolumn}
\usepackage{xcolor}
\usepackage{bm}
\usepackage{soul}

\begin{document}

\title{Singular Spectrum Analysis of Two Photon Interference from Distinct Quantum Emitters}
\author{Rocco Duquennoy}
\affiliation{Dipartimento di Fisica, Universit\`a degli Studi di Napoli, via Cinthia 21, Fuorigrotta 80126, Italy}
\affiliation{Istituto Nazionale di Ottica - CNR  (CNR-INO), Via Nello Carrara 1, Sesto F.no 50019, Italy}

\author{Maja Colautti} 
\author{Pietro Lombardi} 
\affiliation{Istituto Nazionale di Ottica - CNR  (CNR-INO), Via Nello Carrara 1, Sesto F.no 50019, Italy}
\affiliation{European Laboratory for Non-Linear Spectroscopy (LENS), Via Nello Carrara 1, Sesto F.no 50019, Italy}

\author{Vincenzo Berardi}
\affiliation{Dipartimento Interateneo di Fisica, Politecnico di Bari, Via Orabona 4, 70126 Bari, Italy}

\author{Ilaria Gianani}
\affiliation{Dipartimento di Scienze, Universit\`a degli Studi Roma Tre, Via della Vasca Navale 84, 00146 Rome, Italy}

\author{Costanza Toninelli}
\affiliation{Istituto Nazionale di Ottica - CNR  (CNR-INO), Via Nello Carrara 1, Sesto F.no 50019, Italy}

\affiliation{European Laboratory for Non-Linear Spectroscopy (LENS), Via Nello Carrara 1, Sesto F.no 50019, Italy}

\author{Marco Barbieri}
\affiliation{Dipartimento di Scienze, Universit\`a degli Studi Roma Tre, Via della Vasca Navale 84, 00146 Rome, Italy}
\affiliation{Istituto Nazionale di Ottica - CNR  (CNR-INO), Via Nello Carrara 1, Sesto F.no 50019, Italy}

\begin{abstract}
Two-photon interference underlies the functioning of many quantum photonics devices. It also serves as the prominent tool for testing the indistinguishability of distinct photons. However, as their time-spectral profile becomes more involved, extracting relevant parameters, foremost the central frequency difference, may start suffering difficulties. In a parametric approach, these arise from the need for an exhaustive model combined with limited count statistics. Here we discuss a solution to curtail these effects on the evaluation of frequency separation relying on a semiparametric method. The time trace of the quantum interference pattern of two photons from two independent solid-state emitters is preprocessed by means of singular spectral analysis before inspecting its spectral content. This approach allows to single out the relevant oscillations from both the envelope and the noise, without resorting to fitting. This opens the way for robust and efficient on-line monitoring of quantum emitters.
\end{abstract}

\maketitle


\section{\label{sec:introduction} Introduction}

 The inspection of physical phenomena often confronts experimentalists with the challenge of monitoring a system whose evolution is affected by concurrent or even competing events with different characteristic times. 
This happens, for instance, for quantum light sources in the solid state: while a perfectly isolated two-level system yields a single photon source fully defined by the lifetime of the excited state, real-world systems embedded in a host material are subject to numerous sources of decoherence acting at distinct time scales, including charge and spin fluctuations, coupling to the phonon bath, as well as the presence of defect dynamics in the host material~\cite{Maier_1996_two,Beyler_2013_direct,Wolters_2013_measurement,Kuhlmann2013_charge}. Typical consequences are the broadening of the emitter optical transition with respect to the ideal Fourier limit and the emergence of spectral fluctuations, also known as spectral diffusion. While the details of the microscopic origin can vary, these are ubiquitous phenomena in systems like color centers in diamond~\cite{Jantzen2016}, semiconductor nanorods~\cite{Mueller_2004}, carbon nanotubes~\cite{Jeantet2018_interplay}, quantum emitters in hexagonal boron nitride~\cite{Spokoyny2020,White:21}, molecules~\cite{Reilly1993,Shkarin2021,Gmeiner2016}, and quantum dots~\cite{Vural_2020,Kuhlmann2013_charge,Lyasota2019}. It follows that resolving the complete photo-physical characterization of solid-state single-photon sources and, especially, capturing the dynamic evolution of their spectral properties is a hard task. It typically requires the estimation of several interplaying parameters via independent measurements~\cite{Duquennoy:22}, which is not only time-consuming, but also limited by by the measurement time scale (usually minutes) during which the single system parameters can vary accordingly to faster decoherence dynamics occurring with different time scales.

\begin{figure}[h!]
\centering
\includegraphics[width=0.9\columnwidth]{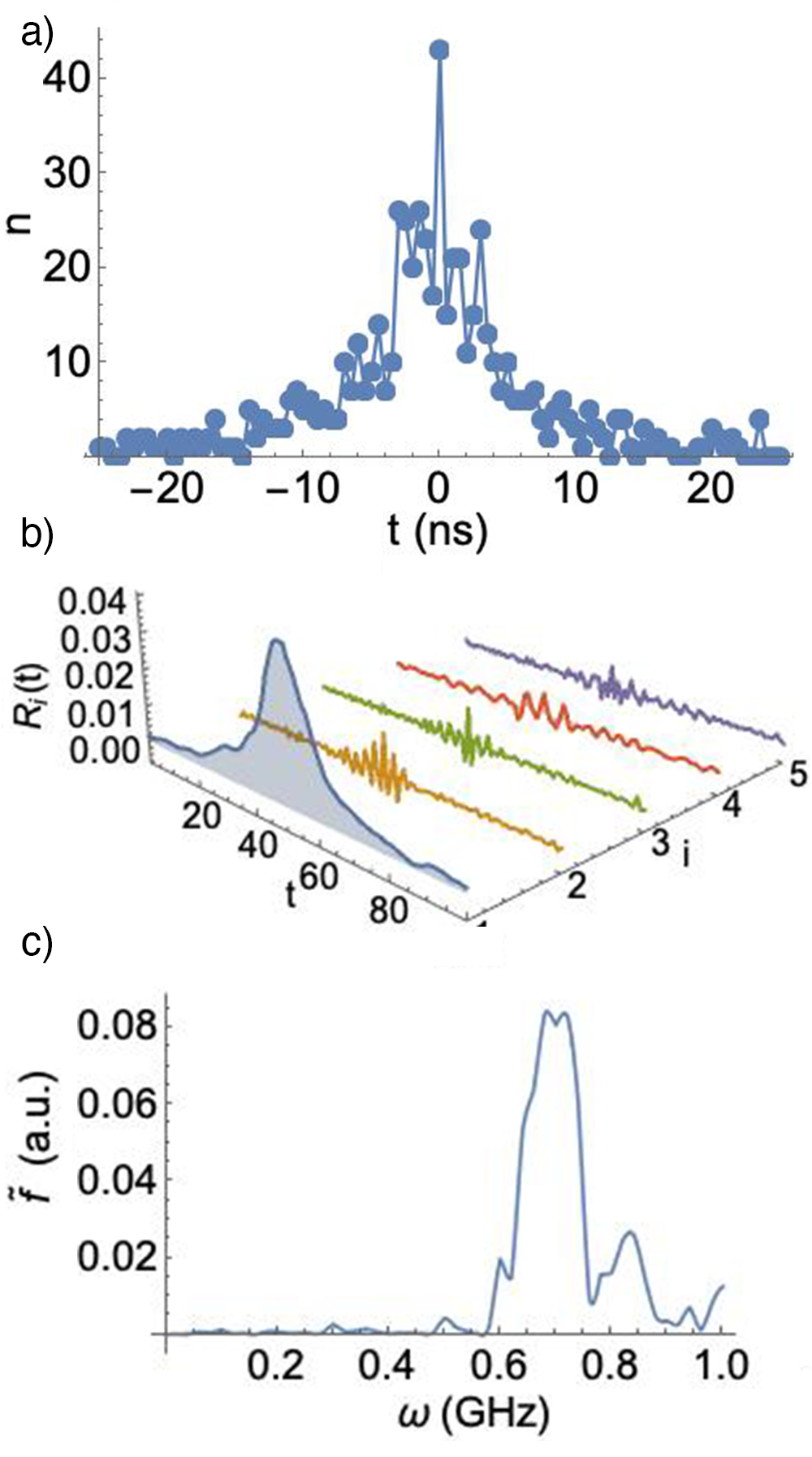}
\caption{\textcolor{black}{SSA analysis of a TPI time trace. Data are collected as coincidence counts in (a), with a delay $t$ monitored by time-tagging the events. The measurement time bin is 0.5ns, while the overall counting time is 60s. Principal component analysis is carried out to deliver the $A_i$'s by means of \eqref{eq:PrinC} and, from those, the $R_i$'s by means of \eqref{eq:RicC} (b.). The Fourier transform $\tilde f$ of the filtered signal is taken (c.) and the value corresponding to its highest peak is considered as the estimated $\delta \nu$.}}
\label{fig:simulatedHOM}
\end{figure}

Tackling the problem with the tools of multiparameter estimation~\cite{PhysRevLett.72.3439, PhysRevLett.96.010401,lilMagda,Demkowicz_Dobrza_ski_2020,ALBARELLI2020126311} offers an elegant solution. However, in its standard setting, a statistical model is needed to describe how the detection probabilities depend on the system parameters. This demand a high level of knowledge about the whole process which may not be available, or could be compromised as the measurement progresses~\cite{PhysRevA.99.053817}. Interestingly, there exists the possibility of estimating key system features also in the absence of a statistical model, applying semiparametric methods~\cite{PhysRevResearch.1.033006,PhysRevX.10.031023,PhysRevA.104.L061701,PhysRevA.105.012411}. In this approach, the values of the parameters are built by means of a more direct manipulation of the experimental data based on the extraction of the principal oscillatory components, rather than relying on fitting routines as for maximum likelihood estimators, or on expected probabilities as for the Bayesian case~\cite{RevModPhys.90.035005,9049135}. This allows establishing an asymmetry between parameters of interest and those acting as nuisance~\cite{Suzuki_2020}, a distinction that is less pronounced in standard approaches.
Indeed, the experience of other fields in dealing with the inspection of complex systems, like climatology~\cite{Ghil_2002}, shows that isolating oscillatory components in time series provides a more compelling evidence than elaborating on fits. To this end, however, a simple approach relying on Fourier transforming has the drawback of treating genuine and spurious effect on the same pace.  In other words, even isolating fast AC components of the transform would not lead to physical insight, since signal and noise would be present.  

In this work, we  show how the application of a semiparametric method based on singular spectrum analysis (SSA) can resolve the problem of analysing multi-scale dynamics in solid-state quantum emitters. This employs data itself to determine which are the most informative features of the time trace of a two-photon interference (TPI) profile (for a pedagogical guide see~\cite{Claessen}). 
In particular, we report on the application of SSA for the determination of the frequency separation of two molecule-based single-photon sources from their Hong-Ou-Mandel two-photon interference (TPI) profile \cite{PhysRevLett.59.2044}. Besides standing as a key enabling process for many photonic protocols for quantum technologies~\cite{RevModPhys.79.135,Bouchard_2020}, TPI is an exquisite probe for the level of distinguishability~\cite{aggie,PhysRevLett.96.240502,Shields,PhysRevLett.100.133601,PhysRevLett.123.080401}, being extremely sensitive to the frequency detuning of the photon pair. At the same time, for solid-state emitters, a faithful extraction of the spectral information from its profile is challenged by the complexity of a multiparameter analysis and by a signal-to-noise-ratio (SNR) that is typically low. In this sense, the TPI profile is particularly suitable to explore the potentials of SSA in the domain of quantum technologies. Its shape intrinsically depends on multiple system parameters, related not only to the emitters pair but also to the setup configuration. In addition, we can leverage on an accurate model of the phenomenon to test the estimation task via simulations. In particular, we will show that SSA provides a powerful means to extract the spectral difference of two single-photon streams in a frequency range of tens of GHz, also in prohibitive SNR regimes in which standard fitting algorithms would be inconclusive. Differently from a multiparameter fitting routine, our method requires no other independent measurement, nor any assumption on the model. This enables a time-effective analysis of the joint spectral properties of the photon pair. \textcolor{black}{Via the processing of multiple interference profiles measured at different times and separated by only 60 seconds, we can resolve the spectral dynamics of the photon pair. This is a promising starting point towards real-time diagnostics.}

\section{Results}

\subsection{TPI from independent solid-state photon sources}

\begin{figure}[t]
     \centering
    \includegraphics[width=1\columnwidth]{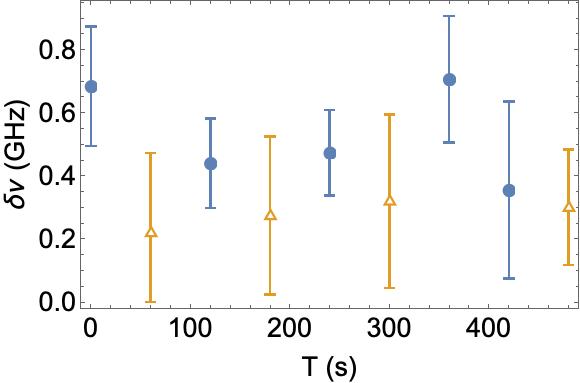}
    \caption{Results on a continuously running experiment. The points refer to measurements acquired in 60s yielding to $n_{\rm tot}\simeq500$ events. The full circles are obtained by including both 2nd and 3rd component in the Fourier analysis, the open triangles are obtained using the 2nd component only - see Appendix.}
    \label{fig:experiment}
\end{figure}

The following analysis makes use of the experimental data appearing in~\cite{Duquennoy:22}, for which TPI has been achieved between photons from distinct molecular emitters on the same chip and brought to resonance via a recently demonstrated laser-induced tuning technique~\cite{Colautti_2020_laser}. Specifically, the quantum emitters consist in single molecules of dibenzoterrylene (DBT) embedded in anthracene nanocrystals~\cite{Pazzagli_2018_self}, which have proven to be excellent quantum light sources~\cite{lombardi2020molecule}, showing bright single-photon emission of high purity even at room temperature~\cite{murtaza2022}. At cryogenic temperature, they exhibit a narrow zero-phonon line (ZPL) of few tens of MHz and emit highly indistinguishable photons, as assessed in recent TPI demonstrations~\cite{Lombardi2021_triggered, Duquennoy:22}. At the same time, with such a narrow optical transition, single-molecules are uniquely sensitive probes of their local environment, and this directly impacts their TPI profile.

More in detail, referring to the experiment in~\cite{Duquennoy:22}, we consider two photons originating from distinct sources operating at central frequencies $\nu$ and $\nu+\delta\nu$. These are made to impinge on a beam splitter with reflectivity $1/2$ and with a relative delay $t$. Hence, two single-photon detectors at the distinct outputs are used to measure the photon arrival times and reconstruct the coincidences. 
According to an adaptation~\cite{Lombardi2021_triggered,Duquennoy:22} of the model in~\cite{Kambs_2018} for photons from distinct emitters, the interferogram observed in the histogram of coincidences around zero time delay is expected to follow a curve described by
\begin{equation}
\begin{aligned}
\label{eq:kambs}
    g^{(2)}_{HOM}(t)=&\frac{1}{4(\tau_1+\tau_2)}\left(e^{-|t|/\tau_1}+e^{-|t|/\tau_2}\right)+\\
    &-\frac{v}{2(\tau_1+\tau_2)}e^{-\gamma |t|-2\pi^2\Sigma^2t^2}\cos(2\pi\, \delta\nu \,t).
    \end{aligned}
\end{equation}
We can here recognise a first term associated to the photons wave-packets, with $\tau_1$ ($\tau_2$) being the time constant in the exponential envelope of the first (second) photon. This also describes what is observed when interfering fully distinguishable photons. The second term accounts for quantum beats caused by the frequency separation $\delta \nu$ and the delay $t$~\cite{lettow10b}, \textcolor{black}{similar to those observed in~\cite{PhysRevLett.61.54,PhysRevA.50.2564,PhysRevLett.93.070503,PhysRevA.85.021803}. }
This second term also contains decoherence effects acting at different time scales on the molecule transition and arising from its coupling to the environment. In particular, the dephasing rates $\Gamma_i$ are included in $\gamma=1/(2\tau_1)+1/(2\tau_2)+\Gamma_1+\Gamma_2$, whereas spectral wandering effects are described by $\Sigma^2=\sigma_1^2+\sigma_2^2$, which is the sum of the variances of each molecule central emission frequency over the measurement acquisition time. The phenomenological parameter $v$, also called $v$-factor, identifies an effective quantum interference visibility accounting for the deviations from ideal conditions. These include the beam splitter reflectivity departing from $1/2$ and the finite integration time, which corresponds to a bandpass filtering in the spectral domain. In Fig.\ref{fig:simulatedHOM} a, we show a typical measurement of the interference profile from our distinct DBT sources, acquired under pulsed excitation. Each data point $n(t)$ corresponds to the recorded coincidence counts as a function of the time delay and is affected by Poissonian fluctuations. The \textcolor{black}{$t$-axis range} is restricted around the suppressed central peak at zero delay, which contains the spectral information we want to extrapolate.


\subsection{Singular spectral analysis}

\textcolor{black}{The SSA method demands to} arrange the experimental data as a time series $X$ of $L$ values $n(t)$. This series is hence used to construct a matrix $Y$, called `embedded time series' in the literature~\cite{Ghil_2002}, composed as follows. $Y$ has $N_c$ rows, with $N_c$ arbitrarily chosen, \textcolor{black}{the generic $i^{th}$ row is a copy of $X$ delayed by ${i-1}$ time positions and restricted to a length $L-N_c$ \footnote{Alternatively, the length can be kept at $L$, but this would require padding the missing elements with zeros}}. The resulting matrix $Y$ is a large-dimensional system, and the purpose of principal component analysis is to find a reduced space carrying most of its information. We hence define the correlation matrix $C_x=\frac{1}{L}Y.Y^\dag$, with dimension $ N_c\times N_c$, find the eigenvectors $\rho_i$, and finally number them so that the corresponding eigenvalues $\lambda_i$ are in decreasing  order ($|\lambda_1| \ge |\lambda_2| \ge ... \ge |\lambda_{N_{c}}|$) \footnote{This is equivalent to identifying the matrix $U$ in the singular-value decomposition $Y=U\Lambda V^T$}. At this point, we are able to extract the principal components $A_i(t)$ by projecting the original time series over the corresponding eigenvectors $\rho_i$, {\it i.e.} by using the following expression:
\begin{equation}
\label{eq:PrinC}
    A_i(t)=\sum_{j=1}^{N_c}X(t+j-1)\rho_i(j),
\end{equation}
where, for the sake of simplicity, we treat time as a discrete index. From the principal components $A_i(t)$ we can build the reconstructed components $R_i(t)$ of the original signal $X(t)$, defined as
\begin{equation}
\label{eq:RicC}
    R_i(t)=\frac{1}{\mathcal{M}(t)}\sum_{j=\mathcal{L}(t)}^{\mathcal{U}(t)}A_i(t-j+1)\rho_i(j),
\end{equation}
where the limits $\mathcal{L}(t)$ and $\mathcal{U}(t)$, and the normalisation $\mathcal{M}(t)$ depend on the index $t$. Their explicit expressions are directly taken from Ref.~\cite{Ghil_2002} (be aware of a typo in their equation (12) where  $1/M_t$ should be read instead of $M_t$):
\begin{equation}
    \mathcal{M}(t) = 
    \left\{
    \begin{matrix}
         t & 1\leq t \leq N_c-1\\
         N_c  & N_c\leq t \leq  L-N_c+1\\
         L-t+1 & L-N_c+2\leq t \leq L
    \end{matrix}
    \right.
\end{equation}
\begin{equation}
\mathcal{U}(t) = 
    \left\{
    \begin{matrix}
         t & 1\leq t \leq N_c-1\\
         N_c & N_c\leq t \leq L
    \end{matrix}
    \right.
\end{equation}
\begin{equation}
\mathcal{L}(t) = 
    \left\{
    \begin{matrix}
         1 & 1\leq t \leq  L-N_c+1\\
         t-L+N_c & N-N_c+2\leq t \leq L
    \end{matrix}
    \right.
\end{equation}
In particular, time-dependent extremes of the series $\mathcal{U}$ and $\mathcal{L}$ are introduced to build a $R_i(t)$ without losing information at the borders of the domain. Hence, the normalization $\mathcal{M}$ needs to change accordingly.

By construction, summing all the reconstructed components returns the original signal $X(t)$. Furthermore, it follows from the initial decreasing ordering of $\lambda_i$ that the reconstructed components $R_i(t)$ which more importantly impact on the shape of $X(t)$ are the ones with smaller $i$. More specifically, different $R_i(t)$ correspond to different system behaviours, related, for instance, to a general trend of the data, to their oscillation, or to the presence of noise, and an isolated spectral content can be easily extracted by performing the Fourier analysis on a reduced subset of $R_i(t)$ components. According to this procedure, the estimation task is both simpler, with respect to the analysis of the whole time series, and also independent from any model assumption.

In the specific case of our TPI experimental trace in Fig~\ref{fig:simulatedHOM}a, by processing the data by means of SSA we obtain the reconstructed components $R_i(t)$ of Fig.~\ref{fig:simulatedHOM}b. We considered $N_c=5$ as a convenient value to recognize three distinct contributions of $R_i(t)$ to the original signal: the first can be associated to the envelope of the coincidence profile, typically in the first component; the second contains the modulation associated to the spectral properties of the photons pair, generally in the second component. Finally, higher-order terms contain faster modulations.
Following this empirical subdivision, we can isolate the oscillating part of the signal and filter out noise by selecting the second component $R_2(t)$  \textcolor{black}{that contains the most of the spectral information we are after.} The final estimator of $\delta \nu$ is the value of the frequency at the \textcolor{black}{maximum} peak of the Fourier transform $\tilde f(\omega) =\left\vert \mathcal{F}[R_2](\omega)\right\vert^2$ of the filtered signal. \textcolor{black}{We did not rely on fitting procedures or averaging because those methods are more prone to errors due to the residual frequency components in $R_2$ coming from spectral diffusion or noise. An explanatory example is shown in Fig. 1c where multiple peaks are present but there is a clear dominant one. It should be noted that increasing the number of components does not lead to an improved isolation of the beats from the noise. An extended discussion is presented in the Appendix.}

Next, we discuss the application of our method to a continuous monitoring of TPI. Leveraging on the effectiveness of the analysis also in the presence of a low SNR, we could apply the SSA to experimental data acquired over very short time intervals of 60s. 
This enables an almost continuous inspection of the spectral detuning of the photon pair, as shown in Fig.~\ref{fig:experiment}. The value of the total integration time and that of the time binning are chosen according to our experimental parameters, in particular the brightness of our sources and the collection and detection efficiencies, in order to provide a sufficient number of counts.

The observed frequency separation spans over 500MHz,
suggesting that spectral diffusion and dephasing occurred during the measurement interval. These effects have an impact and help explaining the relatively large uncertainties in the Fig.~\ref{fig:experiment}. In particular, these are calculated by first considering the data points $n_i(t)$ to extract new time series of values $n'_i(t)$ from a Poisson distribution of mean $n_i(t)$. Hence, the application of the SSA method to all the new series yields a distribution of $\delta \nu$ whose standard deviation corresponds to the associated uncertainties. We must stress that a best fit procedure to the expression \eqref{eq:kambs} has been attempted on the same data in Fig.~\ref{fig:experiment}, but this has led to inconclusive results since the measurement feature low total counts and long time-scale frequency wandering. A comparison with fitting procedure is shown in the Appendix, thanks to the analysis of Monte Carlo simulations.
The ability to assign a value to $\delta\nu$ even in these extreme SNR conditions vindicates the usefulness of the SSA approach.

\begin{figure}[h!]
 \includegraphics[width=0.9\columnwidth]{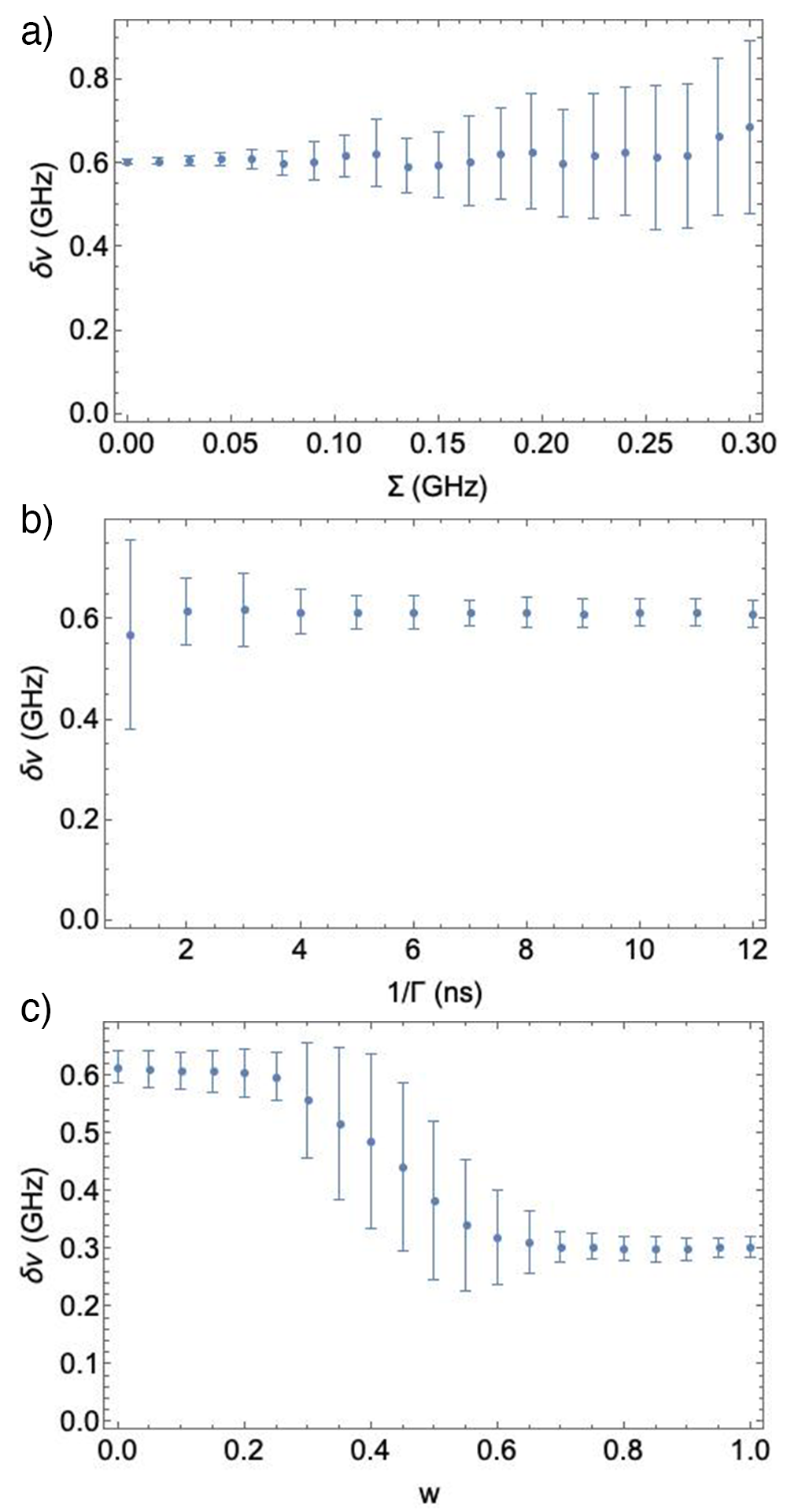}
\caption{Numerical studies of the impact of imperfections. In all panels we consider a TPI profile with $n_{\rm tot}=500$ on average, collected with a resolution $T_{\rm bin}$=0.5ns on an overall delay of 50ns - to be compared with the duration $\tau_1=\tau_2=4.5$ ns. Points and error bars are calculated from 100 Monte Carlo runs as the average and standard deviation over the sample. {(a)} Evaluated frequency separation $\delta\nu$ as a function of spectral diffusion $\Sigma$ with $\delta \nu=0.6$ GHz, $\Gamma = 0.1$ GHz. {(b)} Evaluated frequency separation $\delta\nu$ as a function of \textcolor{black}{the dephasing time} $1/\Gamma$ with $\delta \nu=0.6$ GHz, $\Sigma = 0.05$ GHz. {(c)}  Results in the presence of a frequency jump from $\delta\nu=$0.6 GHz to 0.3 GHz, shown as a function of the weight $w$ of the fastest component. The set values are $\Sigma=0.05$ GHz, $\Gamma =0.1$ GHz.}
    \label{fig:numerics}
\end{figure}

For a deeper understanding of the influence of the different dissipative mechanisms on the estimation of $\delta \nu$ we ran different numerical simulations. Starting this time from the model \eqref{eq:kambs}, we have simulated runs of the experiment in which counts in each time bin are affected by Poisson noise. We first focus on the role of the spectral wandering $\Sigma$, presented in Fig.~\ref{fig:numerics}a. As spectral variations increase, thus lowering the visibility, recovering a precise value becomes harder. This is also supported by the study in Fig.~\ref{fig:numerics}b reporting how the extraction of $\delta\nu$ is influenced by the dephasing rate $\Gamma = \Gamma_1+\Gamma_2$, showing that the impact of both forms of dissipation is qualitatively similar.
Finally, to emulate a spectral jump, we consider a variation of the spectral separation from $\delta\nu = 600$ MHz to $\delta\nu = 300$ MHz occurring while data is being acquired. This is simulated by mixing the statistics pertaining to the two values of $\delta \nu$ with weights \textcolor{black}{$w$} for the component at $600$ MHz, \textcolor{black}{used as the $x$-axis in Fig. 3c}, and $(1-w)$ for the other. The weight can  be interpreted as a rescaled acquisition time with a jump in $\delta\nu$ occurring at the start of the measurement. We observe that SSA gives an intermediate value, depending on $w$, with an increase in the uncertainty as the result of the reduced contrast. This suggests that in real cases the most cautious interpretation of $\delta\nu$ is a weighted average over the measurement time, rather than an instantaneous value. In the Appendix, a comparison of the experimental data with simulations is also presented.

\section{Discussion}

A close inspection of the performance of our method from the metrological point of view has also been performed - the details can be found in the Appendix. The study reveals \textcolor{black}{an evident dependence of the expected error on $\delta \nu$. This can be explained by the fact that the time binning imposes a filter on the fastest frequencies that can be observed.}
At the same time, we also observed a bias in the estimation of $\delta\nu$.  We
attribute it to the discretisation of the time profile dictating, in turn, a discrete set of frequencies. Its impact is also present in the small error bars in Fig.~\ref{fig:numerics}a at low $\Sigma$. Indeed, the reduction of uncertainties in the presence of biased estimators is a well-known effect~\cite{9049135}. We have addressed this by using an interpolation in the frequency domain, which may become less effective in the presence of narrow peaks. 
 We conclude that our method represents an efficient tool for on-line monitoring, while it needs further refinement if meant to be adopted as a tool for metrology. 

The application of semiparametric singular spectral analysis is demonstrated as an intriguing solution for the inspection of TPI. \textcolor{black}{The main advantage stems from the fact a model is not needed, sparing us from the need of estimating multiple parameters. This makes the method resource effective, and, with improvement on the photon flux collection, the monitoring rate can realistically be implemented every 10s.} 

Considering more general perspectives, the development of model-independent methods is key to fostering secure schemes for quantum metrology~\cite{PhysRevA.99.022314,PhysRevA.105.L010401}. While the security analysis of quantum channels makes minimal assumptions, standard metrology postulates its full knowledge and control. The inclusion of non-parametric techniques, including SSA, constitutes a middle ground between the two approaches, on which new protocols can be built.

{\it Acknowledgements.} This work for funded by the EC under the FET-OPEN-RIA project STORMYTUNE (G.A. 899587). 

C.T. acknowledges financial support from the PNRR MUR project PE0000023-NQSTI and from the EMPIR programme (project 20FUN05, SEQUME), cofinanced by the Participating States and by the European Union’s Horizon 2020 research and innovation programme. 

M.B. thanks F. Albarelli, V. Giovannetti, M.A.G. Paris and L.L. Sanchez-Soto for discussion. 

\section*{Appendix}

The availability of a model for our experiment allows to assess the performances of the singular spectral analysis via numerical simulations, in correspondence of different values of the main experimental parameters.

We consider simulations based on the expected profile $g^{(2)}_{\rm HOM}(t)$ with parameters as in Table~\ref{tab:para} when not otherwise specified. These are close to the actual experimental conditions. The finite resolution of the detection is included by integrating around $t$ for a time $T_{\rm bin}$:
\begin{equation}
    \tilde g(t)=\int_{t-T_{bin}/2}^{t+T_{bin}/2} g^{(2)}_{\rm HOM}(t')dt'.
    \label{integral}
\end{equation}

The full time profile is taken as the interval $[-T_{\rm meas},T_{\rm meas}]$ symmetric around zero; this then results in a collection of $2T_{\rm meas}/T_{bin}$ normalised points $g_i=\tilde g(t_i)/\sum_i \tilde g(t_i)$. These are used to generate simulated coincidence counts $n_i$ in each bin by multiplying $g_i$ by the number of events $n_{\rm tot}$: $n_i=n_{\rm tot}g_i$. Finally, fluctuations are accounted for by extracting a new value $n'_i$ from a Poisson distribution of mean $n_i$. 

\begin{table}[h!]
    \centering
    \begin{tabular}{|c|c|}
    \hline
    \text{parameter}& \text{value}\\
    \hline
        $\tau_{1,2}$ & 4.5 ns \\
        $\Sigma$ & 20 MHz \\
        $\Gamma_{1,2}$ & 50 MHz\\
        $T_{\rm bin}$ & 0.5 ns\\
        $n_{tot}$ & 500 \\
        $\it{v}$ & 1 \\
        $\delta\nu$ & 0.6 GHz \\
        \hline
    \end{tabular}
    \caption{Standard parameters employed in the numerical simulations.}
    \label{tab:para}
\end{table}

\subsection{Effects of finite statistics}

\begin{figure*}[t!]
\includegraphics[width=0.7\textwidth]{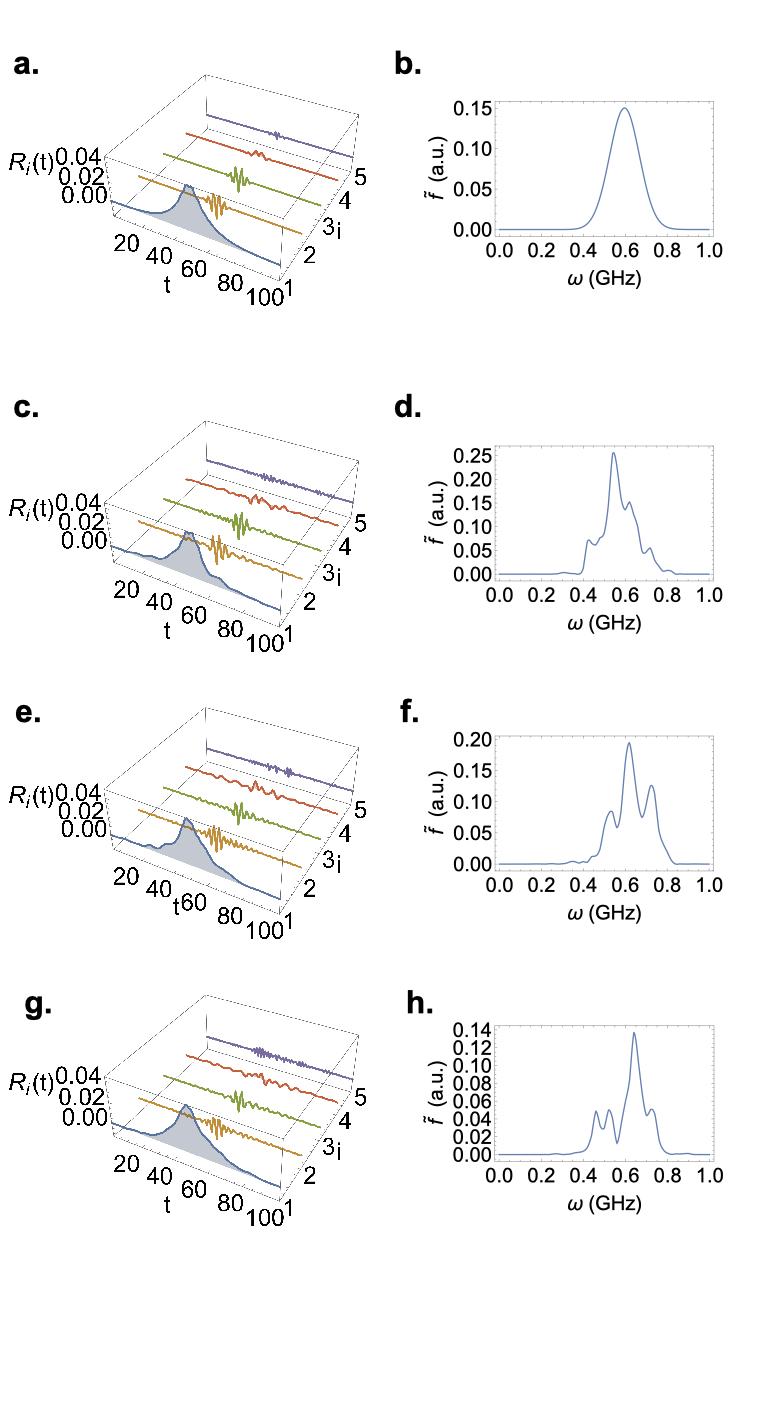}
\caption{a. Ideal reconstructed components with ideal visibility $v=1$ and b. Fourier transform $\tilde f$. The other images show analogue quantities now for simulated reconstruction including Poisson fluctuations with $n_{\rm tot}=500$}
\label{fig:simulation_ideal}
\end{figure*}

We first inspect how the collection of a finite sample affects the spectral content of the reconstructed components \eqref{eq:RicC}.
Figure~\ref{fig:simulation_ideal}a depicts the expected reconstructed components with separation ${\delta\nu = 0.6}$ GHz and perfect contrast ${v=1}$: we observe how the first component broadly describes the envelope, with the modulation appearing in the following ones. In particular, we can inspect the second component $R_2$ as well as the the amplitude squared $\tilde f$ of its Fourier transform, as described in the main text. In this ideal case, obtained applying the method directly to the analytic $\tilde{g}(t)$, the modulation is well isolated, as shown in Fig.~\ref{fig:simulation_ideal}b. When we include Poisson fluctuations in the counts, new modulations appear which affect the spectral content of $\tilde f$. In Fig.~\ref{fig:simulation_ideal}c, e, and g we show instances of simulated profiles affected by such noise at the standard conditions: the envelope is subject to overt distortions, while the higher-order components show different oscillations than before. The analysis of $\tilde f$, however, reveals that the band of the signal still appears around the expected value, although new peaks may appear. 

\begin{figure}[b!]
    \centering
    \includegraphics[width = \columnwidth]{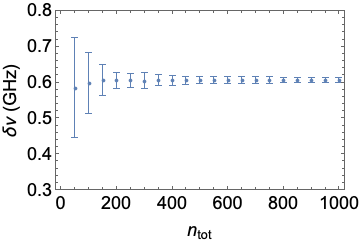}
    \caption{Estimated frequency separation $\delta\nu$ as a function of $n_{tot}$. Each point with its error bars is obtained by 1000 Monte Carlo events.}
    \label{fig:MC_NtotVariation}
\end{figure}

Reducing $n_{tot}$ increases the impact of Poisson noise, thus reducing accuracy and precision in determining $\delta\nu$. A quantitative assessment of this effect is obtained by Monte Carlo simulations, reported in Fig.~\ref{fig:MC_NtotVariation}. We can observe a bias for counts below $n_{tot} =300$, thus being a consequence of low signal along with the increased uncertainty. When a smaller sample is collected, it becomes more likely that one of the spurious peaks observed in Fig.~\ref{fig:simulation_ideal} takes over the one at the real modulation. The Monte Carlo average then shifts towards the center of the allowed frequency interval. This is a general behaviour that can be observed every time an experimental feature washes out the footprint of the beatings in the signal.

\subsection{Effects of a loss of contrast}

\begin{figure}
    \centering
    \includegraphics[width = \columnwidth]{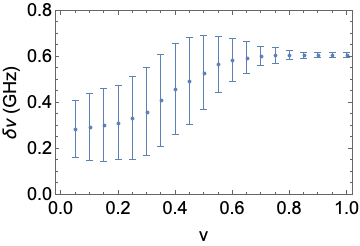}
    \caption{Estimated frequency separation $\delta\nu$ as a function of $v$. Each point with its error bars is obtained by 1000 Monte Carlo events.}
    \label{fig:DNu_vs_vFactor}
\end{figure}

A decrease of the v-factor alters the performance of the method  similarly to a decrease of total coincidence counts. In Fig.~\ref{fig:DNu_vs_vFactor} we present Monte Carlo simulations as a function of $v$ for 500 total coincidence counts. At small $v$, the errors increase and the estimation becomes biased, since the peak in $\tilde f [R_2](\omega)$ decreases in amplitude, eventually becoming comparable to Poisson fluctuations.

\begin{figure*}[t!]
\includegraphics[width=0.7\textwidth]{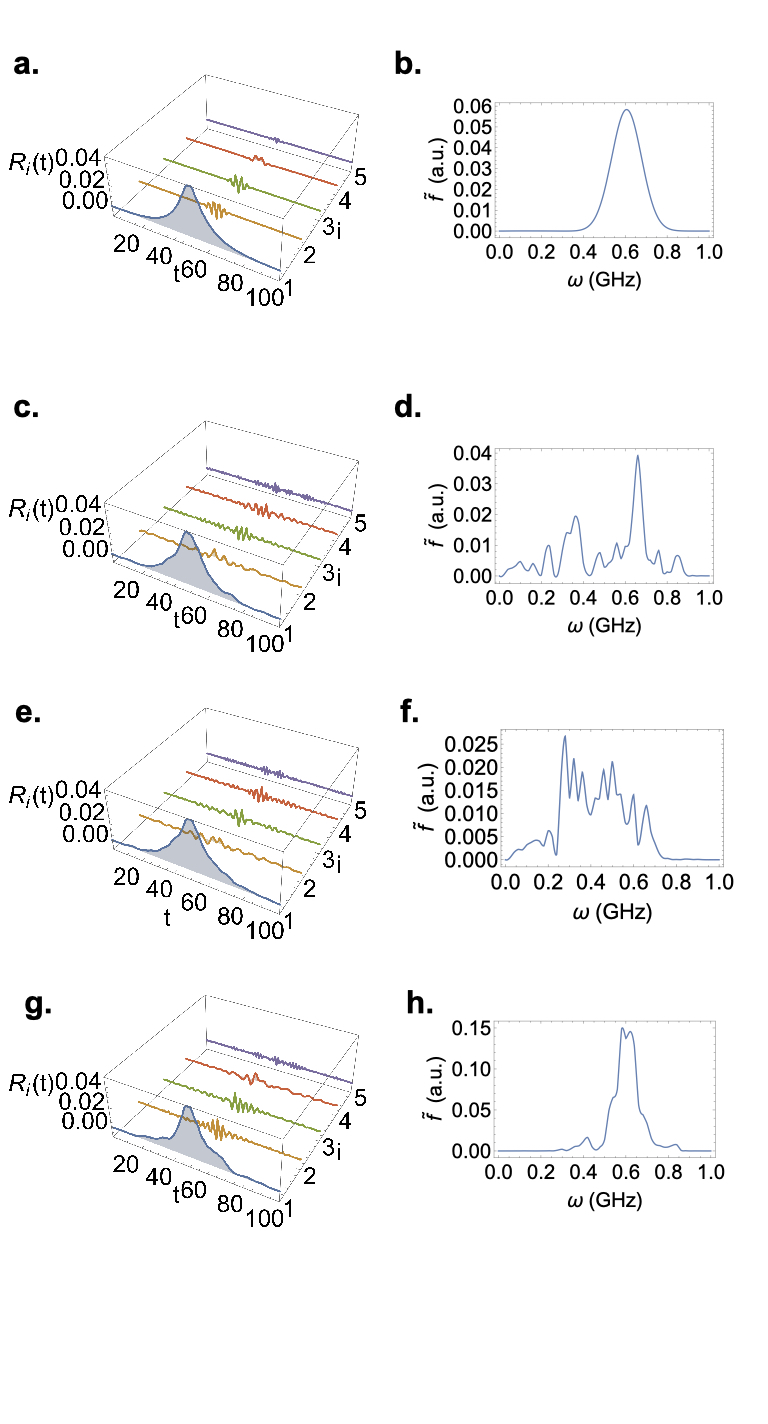}
\caption{a. Ideal reconstructed components with ideal visibility $v=0.65$ and b. Fourier transform $\tilde f$. The other images show analogue quantities now for simulated reconstruction including Poisson fluctuations with $n_{\rm tot}=500$}
\label{fig:simulation_notquite}
\end{figure*}

Here we detail the case for $v=0.65$, shown in Fig.~\ref{fig:simulation_notquite}. For the ideal cases in panels a and b, the modulation depths is reduced, as made evident from the smaller oscillating components. This has a detrimental effect when including the Poisson fluctuations, as their size may be comparable to the expected modulation. As illustrated in Fig.~\ref{fig:simulation_notquite}c-h, these factors may contribute to widen the bandwidth and add features. Nevertheless, a distinct peak is often identified, obtaining a value for $\delta \nu$.

\subsection{The choice of $N_c$}

\begin{figure}
    \centering
    \includegraphics[width =\columnwidth]{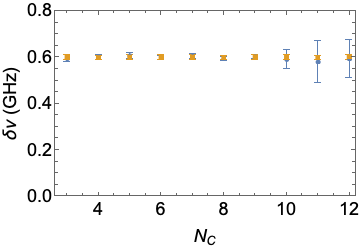}
    \caption{Estimated $\delta \nu$ for varying number of the principal components $N_c$. Blue points refer to the estimation only keeping the $R_2$ components; orange triangles refer to the estimation keeping all components but $R_1$.}
    \label{fig:SSA_vs_Nc}
\end{figure}

Simulations guided our choice of the number of components $N_c = 5$, as well as the restriction to $R_2$ for the spectral analysis. We ran Monte Carlo simulation for different values of $N_c$ reported in Fig.~\ref{fig:SSA_vs_Nc}. The blue points show the results obtained keeping only the second reconstructed component $R_2$ and in orange those obtained by summing up all components except for the first one. In this ideal case with ${v=1}$, the performances are comparable. The increase in the error bars at $N_c = 11, 12$ is ascribed to the fact that, for high enough $N_c$, the information on $\delta\nu$ is spread out on many reconstructed components $R_i$ and thus less robust.

\begin{figure*}[t!]
\includegraphics[width=0.7\textwidth]{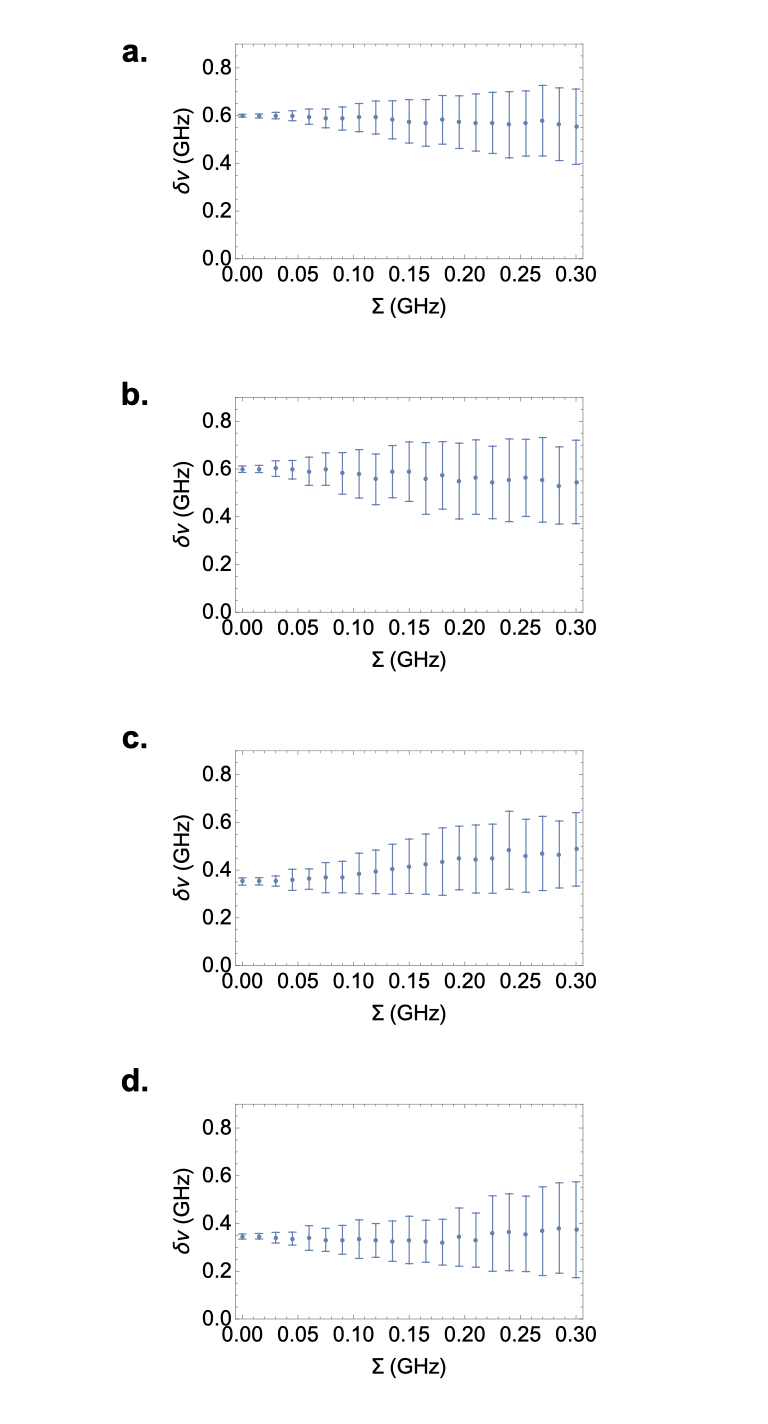}
\caption{ Evaluation of $\delta \nu$ as a function of $\Sigma$. a) set value $\delta \nu = 600$ GHz, v-factor $v=1$; b) set value $\delta \nu = 600$ GHz, v-factor $v=0.6$; c) set value $\delta \nu = 350$ GHz, v-factor $v=0.6$ with limited filtering on the components; d) set value $\delta \nu = 350$ GHz, v-factor $v=0.6$ using the second component.}
\label{fig:simulation_new}
\end{figure*}

The restriction to $R_2$ only finds its justification in the appearance of an artifact at lower $\delta \nu$ at reduced $v$, demanding caution in the selection of the components. Fig.~\ref{fig:simulation_new}a and b report the evaluated $\delta \nu$ {\it vs} the spectral width $\Sigma$ for the set value $\delta \nu = 350$ GHZ at  $v=1$ and $v=0.6$, respectively. For larger values of $\Sigma$ we observe the expected increase of the uncertainties, as well as a tendency to overestimate $\delta \nu$ when components are included according to a looser criterion. We attribute this behaviour to the presence of fast oscillations in the higher components deriving from the Poisson fluctuations. The effect is curtailed when only the second component is included as illustrated in Fig.~\ref{fig:simulation_new}d, although at the cost of larger uncertainties. This dictates the decision of limiting the number of components in the analysis to $R_2$ only.

\subsection{Performance of the best-fit analysis}

A comparison with a standard best-fit procedure has been carried out for a set of 100 simulated data in standard conditions using the {\sc matlab} built-in function $fitnlm$ to the expression in Eq. 1 in the main text, up to an extra multiplicative normalization factor, for a total of 7 free fitting parameters. A Monte Carlo routine proved to be unfeasible since the fit procedure is much slower than the SSA (approximately 1000 times slower without any optimization on either approach). Even when initialising the best-fit parameters to their true values, with $n_{tot}=500$ the routine converged in 75 cases out of 100 to values $\Delta\nu \ge 1$ GHz that cannot be measured with a temporal resolution of $0.5$ ns.

\subsection{Comparison of the simulation with the experimental data}
\begin{figure*}[t!]
\includegraphics[width=0.8\textwidth]{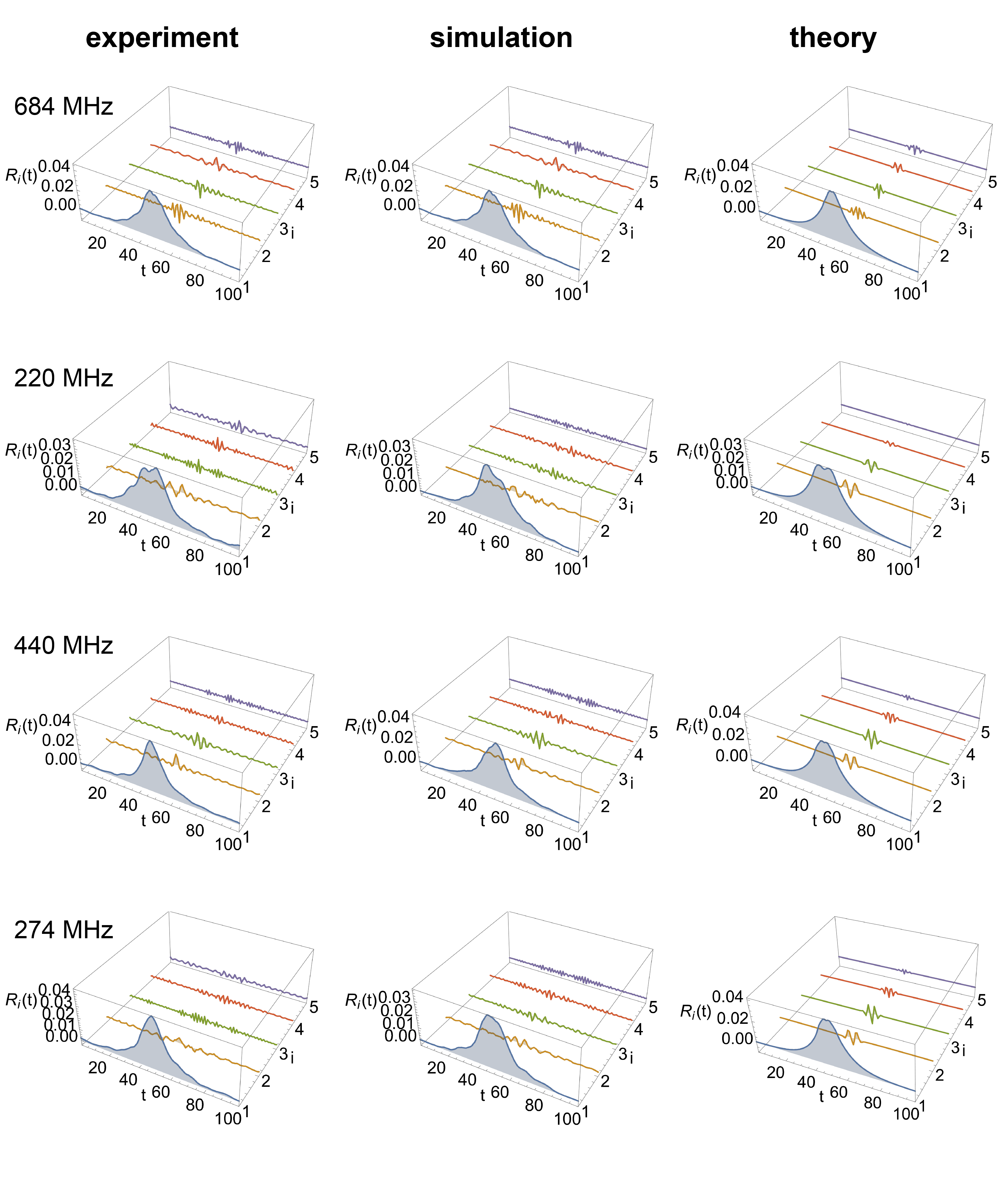}
\caption{Comparison of the experimental reconstructed components with the ones obtained from a simulation and the theoretical ones. }
\label{fig:firstbatch}
\end{figure*}

\begin{figure*}[t!]
\includegraphics[width=0.8\textwidth]{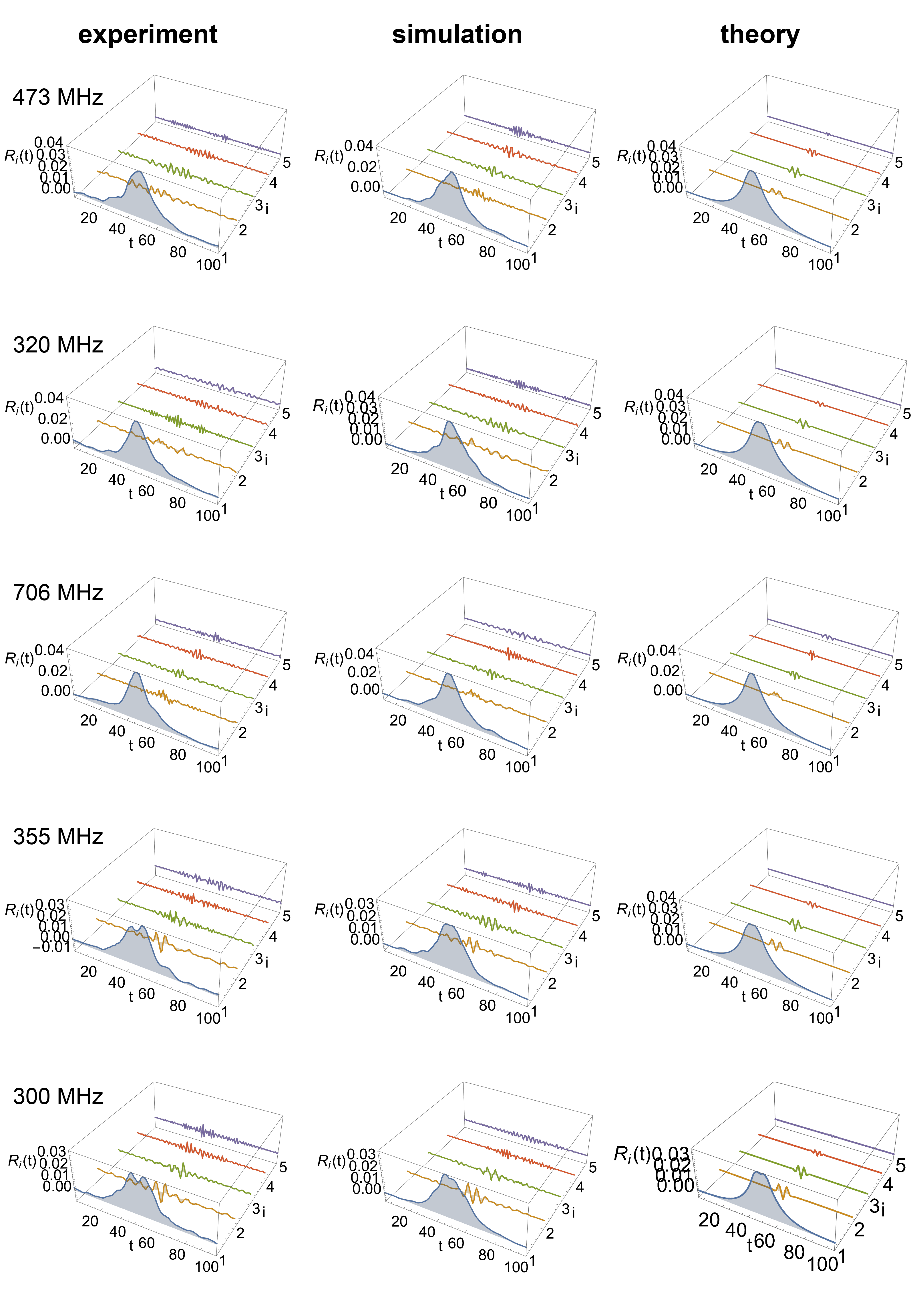}
\caption{Comparison of the experimental reconstructed components with the ones obtained from a simulation and the theoretical ones.}
\label{fig:secondbatch}
\end{figure*}

We can carry out a qualitative comparison between the experimental reconstructed components and those obtained from our model in Figs.~\ref{fig:firstbatch} and~\ref{fig:secondbatch}. The parameters are set according to Table~\ref{tab:para}, with the separation $\delta\nu$ fixed at the estimated value, and the visibility $v$ as a free parameter - the corresponding values are reported in Table~\ref{tab:para2}. The components used for the estimation of $\delta \nu$ are also indicated: preferentially, we have included both components 2 and 3, except when fast oscillations are observed in component 3. Since these occur at frequency close to $1/T_{\rm bin}$ they are likely an artefact from the finite statistics, as corroborated by the simulations presented in Fig.~\ref{fig:simulation_new}.

\begin{table}[b!]
    \centering
    \begin{tabular}{|c|c|c|}
    \hline
    $\delta\nu$\, \text{(MHz)}& v &\text{components}\\
    \hline
        736  & 0.85 & 2\&3\\
        684  & 0.80 & 2\&3\\
        220  & 0.70 & 2\\
        440  & 0.70 & 2\&3\\
        274  & 0.50 & 2\\
        473  & 0.50 & 2\&3\\ 
        320  & 0.50 & 2\\
        706  & 0.50 & 2\&3\\
        355  & 0.60 & 2\&3\\
        300  & 0.70 & 2\\
        \hline
    \end{tabular}
    \caption{Values of visibility used in the simulations, and number of components adopted for frequency estimation}
    \label{tab:para2}
\end{table}

\subsection{Metrological considerations}
\begin{figure}[h!]
\includegraphics[width=\columnwidth]{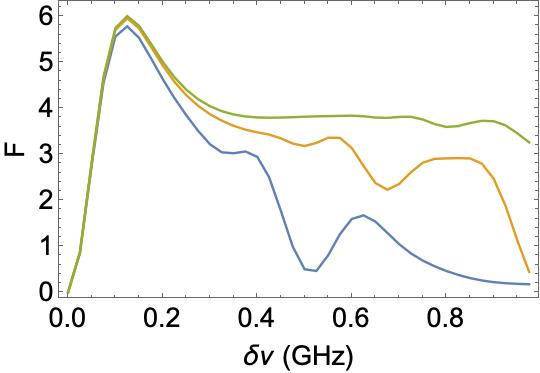}
\includegraphics[width=\columnwidth]{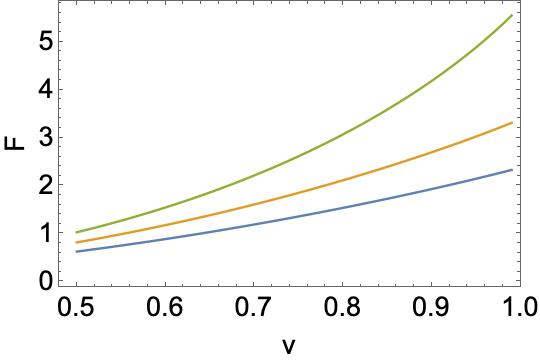}
\caption{Fisher information on $\delta \nu$. The upper panel refers to different values of resolution: from above, the integration time bin is 0.25ns, 0.5ns, and 1ns, the visibility is set to $v=1$, all other parameters are the same as in Table \ref{tab:para}. The lower panel describes the Fisher at different $\delta \nu$ as a function of $v$: from above, the separation is 140 MHz, 440 MHz, 650 MHz, the time bin is 0.5ns, all other parameters are the ones in Table \ref{tab:para}}
\label{fig:fisher}
\end{figure}

Collecting $g_i$ at different discrete times can be considered as a post-selected operation, in which the different values of $t_i$ label the outcomes of a generalised measurement. The associated Fisher information $F(\delta \nu)$ is calculated by its standard expression: 
\begin{equation}
    F(\delta \nu)=\sum_{i} \frac{\left(\partial_{\delta\nu} g_i \right)^2}{g_i}
    \label{fisher}
\end{equation}
This accounts for the fact we only consider the coincidence counts, neglecting the contribution from the individual detectors. The behaviour of $F(\delta \nu)$ is shown in Fig.~\ref{fig:fisher}. The derivatives have been evaluated numerically as a finite ratio, since the integral in Eq. \ref{integral} of the binned distribution prevents to obtain analytical expressions. We observe modulations due to the ratio of $\delta \nu$ to the sampling frequency. Also, $F$ falls to zero as $\delta \nu \rightarrow 0$, as a manifestation of Rayleigh's curse: increasing the resolution does not mitigate this detrimental effect. This occurs in spite of the fact that ours is not a simple measurement of the intensity, but of the second-order coherence. In addition, the information in \eqref{fisher} is a strictly relevant quantity for single-parameter problem: a decrease of the available Fisher information is expected also due to the correlations to the other parameters describing the interference pattern.

\begin{figure}[h!]
\includegraphics[width=\columnwidth]{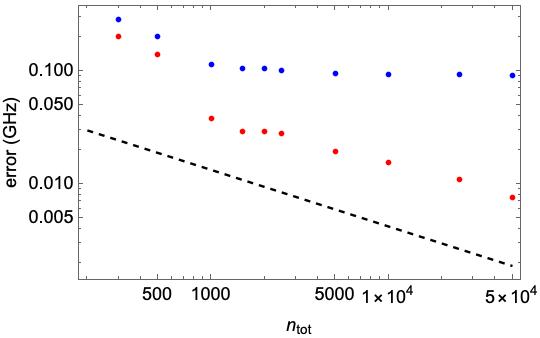}
\includegraphics[width=\columnwidth]{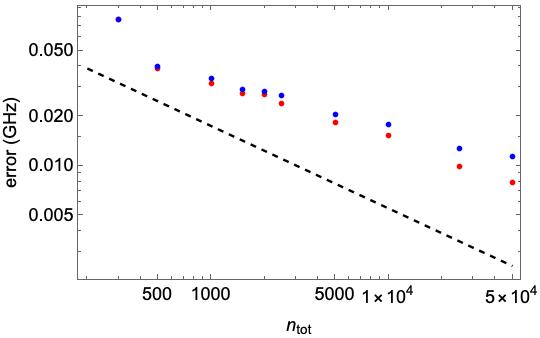}
\includegraphics[width=\columnwidth]{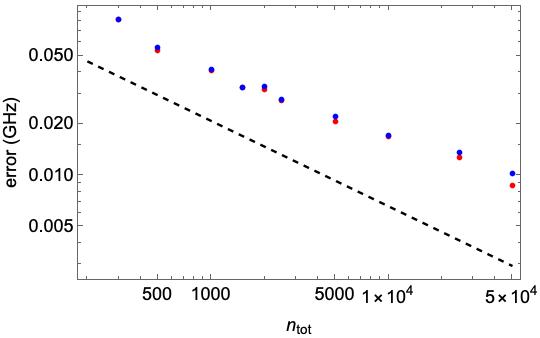}

\caption{Estimation errors as a function of the total number of events $n_{\rm tot}$. The panels refer to $\delta_nu = 140$ MHz, 440 MHz, and 650 MHz, from above. The visibility is set to $v=1$, all other parameters are the same as in Table \ref{tab:para}. In all panels, the blue points are for $\epsilon_{\rm V}$,the red ones for $\epsilon_{\rm RMS}$, the dashed line is the prediction based on the single-parameter Cram\'er-Rao bound.}
\label{fig:biasMC}
\end{figure}

\begin{figure}[h!]
\includegraphics[width=\columnwidth]{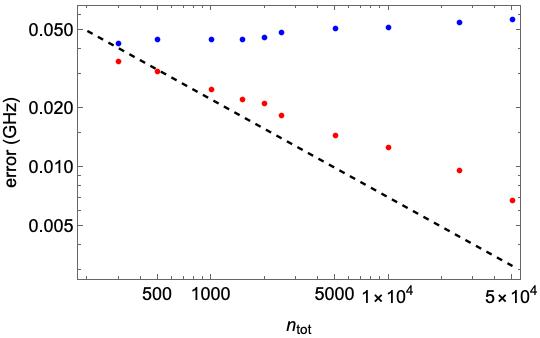}
\includegraphics[width=\columnwidth]{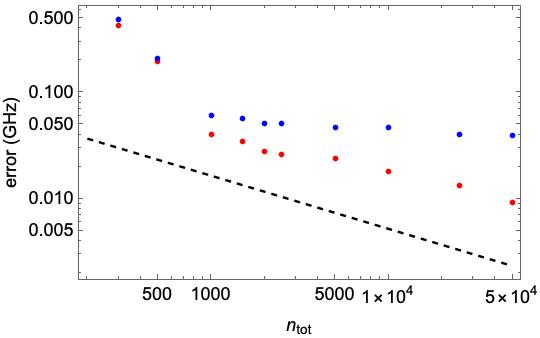}

\caption{Estimation errors as a function of the total number of events $n_{\rm tot}$. The panels refer to $\delta_\nu = 440$ MHz with $T_{\rm bin}=0.25$ns and 1ns, from above.  The visibility is set to $v=1$, all other parameters are the same as in Table \ref{tab:para}, except for $T_{\rm bin}$. In all panels, the blue points are for $\epsilon_{\rm V}$,the red ones for $\epsilon_{\rm RMS}$, the dashed line is the prediction based on the single-parameter Cram\'er-Rao bound.}
\label{fig:biasMC1}
\end{figure}

The expected variability can be estimated by a Monte Carlo routine generating events from a Poisson distribution in each time bin, and proceeding with the SSA. The relevant quantities are the standard deviation of the estimated parameter $\delta\nu$, 
\begin{equation}
    \epsilon_{\rm V} = \sqrt{\langle \left( \delta\nu - \langle \delta\nu\rangle \right)^2\rangle}
\end{equation}
and the root-mean-square departure from the target value $\delta\nu_0$, 
\begin{equation}
    \epsilon_{\rm RMS} = \sqrt{\langle \left( \delta\nu - \delta\nu_0 \right)^2\rangle},
\end{equation}
both calculated over the Monte Carlo sample.
While the former quantifies the variability of the SSA estimator, the latter captures its bias. The results are illustrated in Fig.~\ref{fig:biasMC} and Fig.~\ref{fig:biasMC1}: these report the errors that have been evaluated as a function of the total number of events $n_{\rm tot}$. These reveal the presence of a bias in the estimator: this is due to the fact that frequency is a discretised, although the effect is partly mitigated by the interpolation. Improving the time resolution does not necessarily leads to a corresponding improvement of the bias, see Fig.~\ref{fig:biasMC1}: in fact, the separation of the discrete frequencies is dictated by the time span of the time series.

\bibliography{frequencysep.bib}

\end{document}